\documentclass[11pt,a4paper]{article}
\usepackage{jheppub}
\def\hD{\hat{D}}
\def\dt{\delta\theta}
\def\hpsi{\hat{\psi}}
\def\hdel{\hat{\delta}}
\def\hphi{\hat{\phi}}
\title{Novel Ambiguities in the Seiberg-Witten Map and the Emergent Gravity}
\author{B.~Muthukumar} 
\affiliation{Department of Physics, Pondicherry University,\\ 
             R.V.Nagar, Kalapet, Puducherry-605 014, India.}
\emailAdd{muthukumar.phy@pondiuni.edu.in} 
\abstract{
A homogeneous part of the Seiberg-Witten gauge equivalence relation for gauge fields can lead to solutions that involve matter fields in such a way that the gauge equivalence and the dimensional and index structures are preserved. In particular, we consider scalar fields coupled to U(1) gauge fields. The matter fields appear non-linearly in the map. As an application, we analyze the implication of this ambiguity to emergent gravity at the first order in noncommutative parameter and show that the new ambiguity restores the possibility of conformal coupling of real scalar density field that is coupled non-minimally to the emergent gravity induced by gauge fields --- a possibility that is strictly not allowed if we consider only the already known ambiguity in the Seiberg-Witten map. 
}
\keywords{Noncommutative Field Theories, Seiberg-Witten map, Emergent Gravity.}
\arxivnumber{1408.5478}
\begin{document}
\maketitle
\flushbottom
%
\section{Introduction}
%
Field theories on noncommutative (NC) Moyal space have many interesting properties. One important aspect is the Seiberg-Witten map. In such spaces even a $\hat{U}(1)$ gauge theory becomes a non-Abelian theory. Although the $\hat{U}(1)$ and the other gauge groups defined on these spaces are fundamentally different from ordinary gauge groups, it turns out that we can map the ordinary gauge fields $A_{\mu}$ to noncommutative gauge fields ${\hat{A}}_{\mu}$ and the map is commonly called Seiberg-Witten (SW) map \cite{Seiberg:1999vs}. Since this mapping is done in a way that preserves the gauge equivalence in both ordinary and noncommutative cases, the map is not unique and there are ambiguities in the relation between $A_{\mu}$ and ${\hat{A}}_{\mu}$ \cite{Asakawa:1999cu}. Such freedom in the SW-map turns out be necessary in the process of renormalization \cite{Bichl:2001cq}. 

Another important property is concerned with the emergent gravity phenomenon which is established using the SW-map. The noncommutative spaces can arise from the union of Heisenberg's uncertainty principle with Einstein's theory of classical gravity \cite{Doplicher:1994tu} and there are efforts to work out the inverse problem of whether the noncommutative theories induced by Moyal product in flat spacetime possesses any signature of gravity. An important step along this direction was taken in \cite{Gross:2000ph} where it was shown that translations along noncommutative directions are equivalent to gauge transformations --- a property similar to the one in general relativity where local translations are gauge transformations associated with general coordinate transformations. The next step was taken in \cite{Rivelles:2002ez} which established at the first order in the noncommutative parameter $\theta$ that the  $\hat{U}(1)$ gauge theory in NC Minkowski spacetime, after SW-mapping, is the same as the ordinary gauge theory coupled to gravitational background and that the emergent gravitational field is generated by ordinary gauge fields. 

The notion of emergent gravity has been extensively studied in the NC scenario and in other contexts \cite{Yang:2004vd,Banerjee:2004rs,Muthukumar:2004wj,Yang:2006dk,Yang:2008fb,Steinacker:2010rh,Lee:2012px,Lee:2012rb,Lee:2012ju,Rivelles:2013ica,Seiberg:2006wf,Carroll:2010zza,Yang:2011bd,Carlip:2012wa}. Recently, it was shown in \cite{Rivelles:2013ica} that the ambiguity terms in the Seiberg-Witten map change the emerging metric in the emergent gravity scenario and that a real NC scalar field is mapped to an ordinary scalar density field which is coupled non-minimally to the emerging gravity. It was also shown that in such cases the conformal coupling is strictly not allowed. This is in deep contrast to other metric theories of gravity in which the non-minimal coupling of scalar field of non-gravitational origin has conformal coupling to gravity because of the Einstein equivalence principle \cite{Chernikov:1968zm,Sonego:1993fw,Grib:1995xm,Grib:1995xp,Faraoni:1996rf,Faraoni:1998qx}.

In this paper, we show that in addition to the ambiguities in the SW-map reported so far \cite{Asakawa:1999cu,Suo:2001ih}, there are further ambiguities in the map which can arise in the presence of matter fields. We also show that the inclusion of these novel ambiguities in the context of emergent gravity restores the possibility of conformal coupling of scalar density field which is  coupled non-minimally to the emerging gravitational field. 

The paper is organized as follows. In Section 2, we briefly review the ideas of Seiberg-Witten map for the gauge and matter fields and the ambiguities in them. We show in Section 3 that the presence of matter fields can lead to further ambiguities in the map and that the NC gauge field in general can also depend on ordinary matter fields. The implications of these ambiguities to emergent gravity is discussed in Section 4. We close with concluding remarks in Section 5.
%
\section{Ambiguities in Seiberg-Witten map}
%
The noncommutative gauge transformations are defined as
\begin{align}
\delta_{\hat{\lambda}}\hat{A}_{\mu} & = \partial_{\mu}\hat{\lambda} + i (\hat{\lambda}*\hat{A}_{\mu}- \hat{A}_{\mu}*\hat{\lambda})  \, ,\\
\delta_{\hat{\lambda}}\hat{F}_{\mu\nu} &= i [\hat{\lambda},\hat{F}_{\mu\nu}]_* \,\, .
\end{align}
One difference between the ordinary and noncommutative gauge theory is that the products of functions are replaced by the Moyal *-products. Another difference is that since a function $\hat{C} = \hat{A} * \hat{B}$ is a function of the noncommutative parameter $\theta$ and since an NC gauge field $\hat{A}_{\mu}$ can always be written as a gauge transformation of another gauge field $\hat{A}'_{\mu}$, the NC gauge field in general depends on $\theta$, and hence we can have a series expansion of $\hat{A}_{\mu}$ in orders of $\theta$. The only consistent way by which we can work out the explicit forms of  $\hat{A}_{\mu}$ at each order of $\theta$ is through the amazing prescription given by Seiberg and Witten \cite{Seiberg:1999vs} and it is  called Seiberg-Witten map. The basic idea is to write the gauge field $\tilde{A}$ in an NC theory with NC parameter $\theta+\delta\theta$ in terms of the gauge field $\hat{A}$ in the theory with NC parameter $\theta$ in a way that preserves the gauge equivalence relation:
\begin{align}
 \tilde{A}(\hat{A}+\hat{\delta}_{\hat{\lambda}}\hat{A}) = \tilde{A}(\hat{A}) + \tilde{\delta}_{\tilde{\lambda}}\tilde{A}(\hat{A}). \label{swAmap}
\end{align}
If $\tilde{A}=\hat{A}+\hat{A}^{1}$ and $\tilde{\lambda}=\hat{\lambda}+\hat{\lambda}^{1}$ to first order in $\delta\theta$, then the above gauge equivalence relation gives the first order equation
\begin{align}
 \hat{A}_{\mu}^{1}({\hat{A}+\delta_{\hat{\lambda}}\hat{A}})-\hat{A}_{\mu}^{1}({\hat{A}})-\partial_{\mu}\hat{\lambda}^{1}-i [\hat{\lambda}^{1},\hat{A}_{\mu}]_* -i [\hat{\lambda},\hat{A}_{\mu}^{1}]_*=- \frac{1}{2} \delta\theta^{\alpha\beta}\{\partial_{\alpha}\hat{\lambda} , \partial_{\beta}\hat{A}_{\mu} \}{}_{*}, \label{fosw}
\end{align}
where $\{\}_*$ denotes the anti-commutator. Since the equation (\ref{fosw}) involves two unknowns $\hat{A}^{1}$ and $\hat{\lambda}^{1}$, the solution in general is not unique and there are ambiguities in the map $\tilde{A}=\tilde{A}(\hat{A})$ and $\tilde{\lambda}= \tilde{\lambda}(\hat{\lambda})$ \cite{Asakawa:1999cu}. In particular, the solution to (\ref{fosw}) is given by
\begin{align}
\hat{A}^1 &= -\frac{1}{4}\delta\theta^{\alpha\beta}\{{\hat{A}}_{\alpha},\partial_{\beta}{\hat{A}}_{\mu}+{\hat{F}}_{\beta\mu}\}_*+\alpha\delta\theta^{\alpha\beta}\hD_{\mu}{\hat{F}}_{\alpha\beta} +\beta\delta\theta^{\alpha\beta}\hD_{\mu}[{\hat{A}}_{\alpha},{\hat{A}}_{\beta}]_*, \label{litsolA}\\
\hat{\lambda}^1 &= \frac{1}{4}\delta\theta^{\alpha\beta}\{\partial_{\alpha}{\hat{\lambda}},{\hat{A}}_{\beta}\}_*+2\beta\delta\theta^{\alpha\beta}[\partial_{\alpha}{\hat{\lambda}},{\hat{A}}_{\beta}]_*,
\end{align}
where the terms involving the arbitrary constants $\alpha$ and $\beta$ are the solutions to the homogeneous part of the equation (\ref{fosw}). In addition to these ambiguities, there are further ambiguities in the map which depend on the paths taken in the $\theta$ space to go from one $\theta$-value to another \cite{Asakawa:1999cu}. But these ambiguities can be removed by a combination of gauge transformations, and field redefinitions that would change the functional form of the action \cite{Asakawa:1999cu}. 

Although the original prescription by SW holds only for $\hat{U}(N)$ theories, by allowing the gauge fields and the gauge parameters to be enveloping algebra-valued, we can construct the analogue of SW-map for arbitrary non-Abelian gauge theories \cite{Jurco:2001rq}. 

In the presence of matter fields, we have the following gauge equivalence relation for the matter fields \cite{Jurco:2001rq}:
\begin{align}
\tilde{\psi}(\hat{\psi}+\hat{\delta}_{\hat{\lambda}}\hat{\psi},\hat{A}+\hat{\delta}_{\hat{\lambda}}\hat{A})=\tilde{\psi}(\hat{\psi},\hat{A})+ \tilde{\delta}_{\tilde{\lambda}}\tilde{\psi}(\hat{\psi},\hat{A}) \label{mfSW}.
\end{align}
If  $\tilde{\psi}=\hat{\psi}+\hat{\psi}^{1}$, then for the matter field in the fundamental representation of the gauge group, the above relation in the first order in $\delta\theta$ can be written as
\begin{align}
 \hat{\psi}^{1}(\hat{\psi}+\hat{\delta}_{\hat{\lambda}}\hat{\psi},{\hat{A}+\delta_{\hat{\lambda}}\hat{A}})-\hat{\psi}^{1}(\hat{\psi},{\hat{A}})-i \hat{\lambda}^{1}*\hat{\psi} -i \hat{\lambda}*\hat{\psi}^{1}=- \frac{1}{2} \delta\theta^{\alpha\beta}\partial_{\alpha}\hat{\lambda} * \partial_{\beta}\hat{\psi}. \label{fomsw}
\end{align}
and the solution to (\ref{fomsw}) is given by \cite{Suo:2001ih}
\begin{align}
 \hat{\psi}^1 = i \alpha \delta\theta^{\alpha\beta} {\hat{F}}_{\alpha\beta}*\hat{\psi} + i \beta \delta\theta^{\alpha\beta}[{\hat{A}}_{\alpha},{\hat{A}}_{\beta}]_* * \hat{\psi}. \label{litsolPsi}
\end{align}
Although the above solution is like a gauge transformation, it turns out that the ambiguity due to the non-equivalence of two paths in the $\theta$-space are not completely removed in this case since it involves a gauge parameter different from the one required to remove the ambiguity in ${\hat{A}}^{1}$ due to the non-equivalence of the paths \cite{Suo:2001ih}. 

Also, it was shown in \cite{Rivelles:2013ica} that the ambiguity in the SW-map could cause a physical effect in the context of emergent gravity giving rise to a different geometry when compared with the geometry in the absence of ambiguity. 

In the next Section, we show that  if we relax the condition that NC gauge fields depend only on ordinary gauge fields and their derivatives, then the gauge equivalence relation (\ref{swAmap}) can lead to further ambiguities in the SW-map (\ref{litsolA}) which are different from the above ambiguities. They arise in the presence of matter fields. 
%
\section{More Ambiguities in the presence of Matter fields}
%
In finding the solution to (\ref{fosw}), Seiberg and Witten considered $\tilde{\lambda}$ as a function of ${\hat{\lambda}}$ and ${\hat{A}}$, and $\tilde{A}$ as a function of ${\hat{A}}$. The dimensional constraints that $\delta\theta$ has  power-counting dimension $-2$ and ${\hat{A}}$ and $\partial/\partial x$ have dimension one lead to an expansion of $\tilde{A}$ in powers of ${\hat{A}}$, $\partial/\partial x$ and $\delta\theta$. 

If we depart from the idea that $\tilde{A}$ depends only on ${\hat{A}}$ and its derivatives, and consider only the gauge equivalence and the dimensional aspects, then the homogeneous part
\begin{align}
 \hat{A}_{\mu}^{1}({\hat{A}+\delta_{\hat{\lambda}}\hat{A}})-\hat{A}_{\mu}^{1}({\hat{A}}) -i [\hat{\lambda},\hat{A}_{\mu}^{1}]_*= 0 \label{foph}
\end{align}
of the equation (\ref{fosw}) can lead to further ambiguities which involve the matter fields. If we are to allow $\tilde{A}$ to be a function of matter field $\hat{\psi}$ as well as ${\hat{A}}$, then the eq.(\ref{foph}) takes the form
\begin{align}
 \hat{A}_{\mu}^{1}({\hat{A}+\delta_{\hat{\lambda}}\hat{A}},\hpsi+\hdel_{{\hat{\lambda}}}\hpsi)-\hat{A}_{\mu}^{1}({\hat{A}},\hpsi) -i [\hat{\lambda},\hat{A}_{\mu}^{1}]_*= 0. \label{foparth}
\end{align}
The $n^{\mathrm{th}}$-order counterpart of the above homogeneous equation is 
\begin{align}
 \hat{A}_{\mu}^{n}({\hat{A}+\delta_{\hat{\lambda}}\hat{A}},\hpsi+\hdel_{{\hat{\lambda}}}\hpsi)-\hat{A}_{\mu}^{n}({\hat{A}},\hpsi) -i [\hat{\lambda},\hat{A}_{\mu}^{n}]_*= 0. \label{nthordparth}
\end{align}
%
\subsection{Scalar Fields}
%
In the case of bosonic fields, the fields have power-counting dimension one and it is possible to make $\tilde{A}$ to be $\hpsi$-dependent without violating the gauge-equivalence condition. We do not consider the fermionic case in this paper except noting that the fermionic matter fields have dimension ${3/2}$ in $4$-$d$ and so it is not possible to construct ${\hat{A}}^{1}$ with only ${\hat{A}}$ and $\hpsi$ and their derivatives at the first order in $\theta$. In particular, we consider the case of scalar fields coupled to $\hat{U}(1)$ gauge fields. 
%
\subsubsection{Adjoint Representation}
%
For a real scalar field $\hphi$ in the adjoint representation of the gauge group, the covariant derivative is defined by
\begin{align}
\hat{D}_{\mu}\hphi= \partial_{\mu}\hphi-i [\hat{A}_{\mu},\hphi]_* \label{adjcoderi}. 
\end{align}
The field and its covariant derivative transform under a gauge transformation as 
\begin{align}
 \hat{\delta}_{\hat{\lambda}}\hphi = i [\hat{\lambda}, \hphi]_*  \, , \qquad \hat{\delta}_{\hat{\lambda}}\hat{D}_{\mu}\hphi = i [\hat{\lambda}, \hat{D}_{\mu}\hphi]_*  \, .
\end{align}
If $\hat{G}_1^{(n)}$ is any polynomial function of $\hphi,~\hat{D}_{\mu}\hphi,~ \hat{D}_{\mu}\hat{D}_{\nu}\hphi \ldots$, then under a gauge transformation 
\begin{align}
 \hat{\delta}_{\hat{\lambda}}\hat{G}_1^{(n)}(\hphi,\hat{D}_{\mu}\hphi,\ldots) = i [\hat{\lambda}, \hat{G}_1^{(n)}(\hphi,\hat{D}_{\mu}\hphi,\ldots)]_* \, ,
\end{align}
which implies that the solution to eq.(\ref{nthordparth}) can be written as
\begin{align}\
 \hat{A}_{\mu}^{(n)} = \hat{G}_1^{(n)}(\hphi,\hat{D}_{\mu}\hphi,\hat{D}_{\mu}\hat{D}_{\nu}\hphi,\ldots; \dt)  + (\hat{G}_1^{(n)}(\hphi,\hat{D}_{\mu}\hphi,\hat{D}_{\mu}\hat{D}_{\nu}\hphi,\ldots; \dt))^{\dagger} \, . \label{novambadjnth}
\end{align}
In particular, the solution to the first order equation (\ref{foparth}) becomes
\begin{align}
 \hat{A}_{\mu}^{(1)}= \delta\theta_{\mu\nu}\left( \eta\, \hphi* (\hD^{\nu}\hphi) + \eta^\dagger \,(\hD^{\nu}\hphi)* \hphi\right) \label{novambadj}. 
\end{align}
where $\eta$ is an arbitrary complex constant and $\eta^\dagger$ is its complex conjugate. Note that we have denoted the solution (\ref{novambadj}) with the superscript in parentheses to distinguish it from (\ref{litsolA}). 

As a result of (\ref{novambadjnth}), the higher order terms in $\tilde{\lambda}$ and $\tilde{\phi}$ can have terms nonlinear in $\hphi$ and therefore even the \emph{free field part} of the NC action for matter field can involve highly non-trivial self-interaction terms, after SW-mapping is done.
%
\subsubsection{Fundamental Representation}
%
In the fundamental representation, a complex scalar field transforms as 
\begin{align}
 \hat{\delta}_{\hat{\lambda}}\hphi = i \hat{\lambda}*\hphi \, , \qquad \hat{\delta}_{\hat{\lambda}}\hphi^\dagger = -i \hphi^\dagger * \hat{\lambda} \, ,
\end{align}
The covariant derivative is defined as $\hat{D}_{\mu}\hphi= \partial_{\mu}\hphi-i \hat{A}_{\mu}*\hphi$. 
The solution to eq.(\ref{foparth}) can then be written as
\begin{align}
 \hat{A}_{\mu}^{(1)}=  \delta\theta_{\mu\nu} \left(\zeta (\hat{D}^{\nu}\hphi)*\hphi^{\dagger}+ \zeta^\dagger \hphi *(\hat{D}^{\nu}\hphi)^{\dagger}\right), \label{novambfund}
\end{align}
where $\zeta$ is an arbitrary complex constant. 

In general we can construct the solution to the $n^{\mathrm{th}}$-order equation (\ref{nthordparth}) in the following way. Let $\hat{E}_i$ be the $i^{\mathrm{th}}$ element in the set $\{\hphi,\hat{D}_{\mu}\hphi, \hat{D}_{\mu}\hat{D}_{\nu}\hphi \ldots\}$  and $\hat{E}_j^{\dagger}$ be $j^{\mathrm{th}}$ element in the set $\{\hphi^\dagger,(\hat{D}_{\mu}\hphi)^{\dagger}, (\hat{D}_{\mu}\hat{D}_{\nu}\hphi)^{\dagger} \ldots\}$. Consider a polynomial function $\hat{G}_2^{(n)}(\hat{E}_i*\hat{E}_j^{\dagger})$ that depends on the star product $\hat{E}_i*\hat{E}_j^{\dagger}$ of pair  of elements --- one from each set. Then the gauge transformation of such a polynomial function becomes
\begin{align}
 \hat{\delta}_{\hat{\lambda}}\hat{G}_2^{(n)}(\hat{E}_i*\hat{E}_j^{\dagger}) = i [\hat{\lambda}, \hat{G}_2^{(n)}(\hat{E}_i*\hat{E}_j^{\dagger})]_* \, ,
\end{align}
and the solution to eq.(\ref{nthordparth}) in this case can be written as
\begin{align}
 \hat{A}_{\mu}^{(n)}=  \hat{G}_2^{(n)} + (\hat{G}_2^{(n)})^{\dagger} \label{novambfundnth} \, .
\end{align}
%
\section{Implications to Emergent Gravity}
%
To analyze the implications of the above ambiguities, we take the case of mapping between U(1) gauge fields in the ordinary and noncommutative theories and especially consider the real scalar field in the adjoint representation. In this case, the eq.(\ref{novambadj}) becomes 
\begin{align}\label{foambi}
 A_{\mu}^{(1)}= 2\gamma \theta_{\mu\nu} \phi (\partial^{\nu}\phi) \, ,
\end{align}
where $\gamma$ is some real constant. The effect of the ambiguities as in (\ref{litsolA}) and (\ref{litsolPsi}) in the context of emergent gravity was analyzed in \cite{Rivelles:2013ica}, and it was shown that if the ambiguity terms are included, then a real NC scalar field is mapped to an ordinary scalar density field and that the ordinary scalar density field needs to be coupled non-minimally to the gravitational background induced by ordinary  gauge fields. We briefly review a few ideas of \cite{Rivelles:2013ica} as they are needed for further treatment of the problem. 

Consider the action for a real NC scalar field $\hat{\phi}$ in the adjoint representation in NC Minkowski spacetime: 
\begin{eqnarray}
	\hat{S}_{\hat{\phi}} = \frac{1}{2} \int d^4x \, \hat{D}^\mu \hat{\phi} * \hat{D}_\mu \hat{\phi}, \label{NCScalAct}
\end{eqnarray}
where the covariant derivative $\hat{D}_\mu$ is defined as in (\ref{adjcoderi}). If we don't include (\ref{foambi}), then maps between NC fields and ordinary fields can be written as 
\begin{align}
 \hat{A}_\mu &= A_\mu - \frac{1}{2} \theta^{\alpha\beta} A_\alpha ( \partial_\beta A_\mu + F_{\beta\mu} ) + \alpha \, \partial_\mu \theta F, \label{folitAabel} \\
 \hat{\phi}  &= \phi - \theta^{\alpha\beta} A_\alpha \partial_\beta \phi + \alpha \, \theta F \, \phi,
\end{align}
where $\theta F =\theta^{\alpha\beta}F_{\alpha\beta}$. Upon substituting the above relations, the action (\ref{NCScalAct}), to first order in $\theta$,   takes the form \cite{Rivelles:2013ica}
\begin{eqnarray}\label{foscalact}
	\hat{S}_{\phi} = \frac{1}{2} \int d^4x \left[ (1 + 2 \alpha \, \theta F) \partial^\mu {\phi} \partial_\mu {\phi} - \alpha {\phi}^2 \square \theta F  - 2 \theta^{\mu\alpha} {F_\alpha}^\nu ( \partial_\mu {\phi} \partial_\nu {\phi} - \frac{\eta_{\mu\nu}}{4}  \partial^\lambda {\phi} \partial_\lambda \phi) \right]. 
\end{eqnarray}
This action is compared with the one for a scalar density field $\phi$ with weight $-\omega$, the weight of $\sqrt{-g}$ being $-1$. This field is taken to be without self-interactions in a weak gravitational background and coupled non-minimally to the curvature scalar. The relevant action is 
\begin{eqnarray}
	S^g_{\phi} = \frac{1}{2} \int d^4x \, (\sqrt{-g})^{2\omega+1} \, g^{\mu\nu} \nabla_\mu \phi \, \nabla_\nu\phi + \frac{1}{2} \xi \int d^4x \, (\sqrt{-g})^{2\omega+1} R \, \phi^2, 
\end{eqnarray}
where $\xi$ is the coupling constant and $\nabla_\mu \phi = \partial_\mu + \omega \Gamma^\nu_{\mu\nu} \phi$ is the covariant derivative of the scalar density of weight $-\omega$. In the linearized limit, the metric  $g_{\mu\nu} = \eta_{\mu\nu} + h_{\mu\nu} + \eta_{\mu\nu} h$, where $h_{\mu\nu}$ is traceless, and the above action becomes \cite{Rivelles:2013ica}
\begin{eqnarray}\label{fograviact}
 S^g_{\phi} = \frac{1}{2} \! \int \!\! d^4x \left[ \left( \left(1 + (1+4\omega) h \right)\eta^{\mu\nu} \!-\! h^{\mu\nu} \right) \partial_{\mu} \phi \partial_\nu \phi + ( 3\xi - 2\omega ) \phi^2  \square h  - \xi \, \phi^2  \partial^\mu \partial^\nu h_{\mu\nu} \right] \! .
\end{eqnarray}
Comparing (\ref{fograviact}) with (\ref{foscalact}), we get 
\begin{align}
 h^{\mu\nu} &= \theta^{\mu\alpha}{F_\alpha}^\nu	+ \theta^{\nu\alpha}{F_\alpha}^\mu + \frac{1}{2} \eta^{\mu\nu} \theta F \, ,\label{fometin}\\
        (1+4\omega) h   &= 2 \alpha \theta F  \, , \label{tracemid1}\\ 
        (3 \xi - 2 \omega) h + \frac{\xi}{2}\theta F & = -\alpha \theta F  \, .\label{coupconsmid}
\end{align}
From (\ref{tracemid1}) and (\ref{coupconsmid}), it is clear that if $\alpha = 0$ and $h=0$, then $\xi=0$ and $\omega$ is arbitrary. If $\alpha =0$ and $h\neq 0$, then $\omega=-1/4$ and $\xi$ becomes arbitrary, but $\xi=-1/6$ is not a consistent solution. 

If $\alpha \neq 0$, then $h \neq 0$ and $\omega \neq -1/4$, and in this case, $\xi$ can be worked out to be
\begin{align}
  \xi  &= -\frac{1}{6+\frac{1+4\omega}{2\alpha}} \, . \label{coupcons}
\end{align}
Therefore, the ambiguity term in (\ref{folitAabel}) forces us to consider non-minimal coupling, and the conformal coupling is strictly not allowed. 

If we include the novel ambiguity (\ref{foambi}), then it turns out that this ambiguity term contributes to the action for the scalar field only through the action for the pure NC gauge fields. The action for the NC $\hat{U}(1)$ gauge theory in the NC Minkowski space is given by
\begin{align}
\label{NC_gauge_action}
\hat{S}_{\hat{A}} = -\frac{1}{4} \int d^4x \,\,\, \hat{F}^{\mu\nu} * 
\hat{F}_{\mu\nu},
\end{align}
where $\hat{F}_{\mu\nu} = \partial_\mu \hat{A}_\nu - \partial_\nu
\hat{A}_\mu - i [ \hat{A}_\mu, \hat{A}_\nu ]_*$.
To first order in $\theta$, we write $\hat{F}= F + F^1 + F^{(1)}$, where $F+F^1$ is constructed from (\ref{folitAabel}), and $F^{(1)}$ is the term that involves $\phi$ due to (\ref{foambi}) and it is worked out to be 
\begin{align}
 F_{\mu\nu}^{(1)} = 2 \gamma \left( \theta_{\nu}^{~\lambda} \partial_{\mu}(\phi \partial_{\lambda}\phi) - \theta_{\mu}^{~\lambda} \partial_{\nu}(\phi \partial_{\lambda}\phi) \right).
\end{align}
Then the part of the action (\ref{NC_gauge_action}) that involves $\phi$ can be shown to be equal to 
\begin{align}
 S^{(1)}_{A}   =  \frac{\gamma}{2} \int d^4 x \phi^2 \partial^{\mu}\partial_{\mu} \theta F \, .\label{foambiact}
\end{align}
Adding (\ref{foscalact}) and (\ref{foambiact}), the total action for the scalar field becomes 
\begin{align} \label{corfoscalact}
\hat{S}_{\phi}^{tot} = \frac{1}{2} \int \! d^4x \! & \left[ (1 + 2 \alpha \, \theta F)  \partial^\mu {\phi} \partial_\mu {\phi}  - 2 \theta^{\mu\alpha} {F_\alpha}^\nu ( \partial_\mu {\phi} \partial_\nu {\phi} - \frac{1}{4} \eta_{\mu\nu} \partial^\lambda {\phi} \partial_\lambda \phi) \right. +  \nonumber \\
 ~ & \left.  \phantom{\frac{1}{4}\!\!\!} \quad  +   (\gamma-\alpha){\phi}^2 \partial^{\mu}\partial_{\mu} \theta F   \right] \, .
\end{align}
Comparing (\ref{corfoscalact}) with (\ref{fograviact}), we get  
\begin{align}
 h^{\mu\nu} &= \theta^{\mu\alpha}{F_\alpha}^\nu	+ \theta^{\nu\alpha}{F_\alpha}^\mu + \frac{1}{2} \eta^{\mu\nu} \theta F \, ,\label{fometinv}\\
             (1+4\omega) h & = 2 \alpha \theta F \, ,\label{tracemid}\\       
(3\xi-2\omega) h   +\frac{\xi}{2} \theta F & = (\gamma -\alpha) \theta F \, .\label{midstep}
\end{align}
We can rewrite eqs.(\ref{tracemid}) and (\ref{midstep}) as 
\begin{align}
       h    &= \frac{-\xi+2\gamma}{1+6\xi} \theta F  \, ,\label{trace2}\\
       \xi  &= -\frac{1}{6} \left( \frac{1- \gamma \left(\frac{1+4\omega}{\alpha}\right)}{1+\frac{1}{12}\left(\frac{1+4\omega}{\alpha}\right)}\right) \, . \label{coupcons2}
\end{align}
The novel ambiguity does not affect $h^{\mu\nu}$, but the trace $h$ and the non-minimal coupling are changed. In particular, it follows from eq.(\ref{coupcons2}) that when $\gamma=-1/12$, we get the conformal coupling and when $\gamma$ is zero we get back eq.(\ref{coupcons}). Also, eqs.(\ref{tracemid}) and (\ref{midstep}) imply that if $\omega\neq -1/4$ and $\alpha=0$, then $h=0$ and in this case the conformal coupling $\xi = 2\gamma$. 

If $\omega= -1/4$, then we have $\alpha=0$. But $\xi$ becomes arbitrary in this case. We can infer from eq.(\ref{midstep}) that in this case also, the conformal coupling is a consistent solution and for this coupling $\gamma$ takes the fixed value $-1/12$, but then $h$ cannot be defined. If $\xi \neq -1/6$ then $h$ is the same as in (\ref{trace2}). 
%
\section{Concluding Remarks}
%
We considered the noncommutative (NC) scalar fields coupled to $\hat{U}(1)$ gauge fields. The ambiguities presented in this paper are the solutions to the homogeneous equation (\ref{nthordparth}) of Seiberg-Witten gauge equivalence relation  for the gauge fields, which can involve the matter fields in the presence of matter fields. The scalar fields appear in the ambiguities (\ref{novambadj}) and (\ref{novambfund}) non-linearly already at the level of first order itself. In the higher order of NC parameter, after Seiberg-Witten-mapping is done, the part of the total NC action that involves the scalar field will have terms with highly non-trivial self-interaction terms, even if NC scalar field does not have any direct self-interaction term before mapping. 

On the emergent gravity side, the inclusion of these novel ambiguities does not spoil the emergent gravity phenomenon and we have shown at the leading order in the NC parameter $\theta$, that the NC scalar field theory coupled to $\hat{U}(1)$ gauge field, after the SW-map is applied, leads to a theory of ordinary scalar density field that is coupled non-minimally to the emerging gravitational field. We have also shown that the theory allows room for conformal coupling which is strictly not allowed if we consider only the already known ambiguity in the map.  In this way, the theory is also in conformity with other metric theories of gravity where the Einstein equivalence principle imply that the non-minimal coupling of scalar field of non-gravitational origin has conformal coupling to gravity \cite{Chernikov:1968zm,Sonego:1993fw,Grib:1995xm,Grib:1995xp,Faraoni:1996rf,Faraoni:1998qx}. Since the conformal coupling is possible only for the potentials $V(\phi)= 0$ or $\lambda \phi^4$, it remains to be seen whether this coupling survives in the higher orders of $\theta$. 
%
\acknowledgments
This work was supported by UGC India through the grant F.No.41-1404/2012(SR).
%

\end{document}